# Thicker amorphous grain boundary complexions reduce plastic strain localization in nanocrystalline Cu-Zr

Esther C. Hessong [1], Nicolo Maria della Ventura [2], Tongjun Niu [3], Daniel S. Gianola [2], Hyosim Kim [4], Nan Li [3], Saryu Fensin [3], Brad L. Boyce [5,6], Timothy J. Rupert [1,7,8,*]

[1] Department of Materials Science and Engineering, University of California, Irvine, CA 92697, USA

[2] Materials Department, University of California, Santa Barbara, Santa Barbara, CA 93106, USA

[3] Center for Integrated Nanotechnologies, Materials Physics Applications Division, Los Alamos National Laboratory, Los Alamos, NM 87545, USA.

[4] Materials Science and Technology Division, Los Alamos National Laboratory, Los Alamos, NM 87545, USA

[5] Center for Integrated Nanotechnologies, Albuquerque, NM 87185, USA

[6] Sandia National Laboratories, Albuquerque, NM 87185, USA

[7] Hopkins Extreme Materials Institute, Johns Hopkins University, Baltimore, MD 21218, USA

[8] Department of Materials Science and Engineering, Johns Hopkins University, Baltimore, MD 21218, USA

* Email: tim.rupert@jhu.edu

**Abstract**

Amorphous grain boundary complexions have been shown to increase the plasticity of nanocrystalline alloys as compared to ordered grain boundaries. Here, the effect of an important structural descriptor, amorphous complexion thickness, on the plasticity and failure modes of nanocrystalline Cu-Zr is studied with *in-situ* compression testing, with over 50 micropillars tested. Two model materials were created that differ only in their complexion thickness, with one having a thicker complexion population than the other. The sample with thinner complexions was more likely to experience non-uniform plastic deformation in the form of localized plastic flow or shear banding. In contrast, the sample with thicker complexions displayed more homogeneous plasticity and higher damage tolerance; thicker amorphous complexions suppress localization by absorbing defects. This work demonstrates that increasing complexion thickness can be beneficial for stable plastic flow in nanocrystalline alloys, by improving resistance to strain localization and premature failure.

Grain refinement to the nanoscale can result in dramatic increases to material strength, particularly in metals [1, 2]. The increase in relative volume fraction causes grain boundaries to play a dominant role in the deformation of nanocrystalline metals. Compared to coarse-grained counterparts where plastic deformation occurs by dislocation glide over relatively large distances, grain boundaries in nanocrystalline materials act not only as barriers to dislocation motion but also become involved in dislocation nucleation and absorption [3, 4]. These events involve intense rearrangements in the boundary and can lead to premature failure, making grain boundary structure of utmost importance. While the grain boundaries of most materials are crystalline with periodic atomic structure, some specially designed materials contain a grain boundary region that is amorphous, possessing no long-range atomic periodicity. Amorphous grain boundary complexions, where a solid yet disordered structure exists, in particular have been shown to improve the compressive strength, plasticity, and toughness of nanocrystalline materials in recent years [5-9]. Amorphous grain boundaries deform through collective rearrangements and can therefore absorb dislocations without cracking as easily as ordered boundaries, which enables greater plastic deformation and plastic strains.

There is a vast body of literature on the deformation mechanisms of fully-amorphous metals such as bulk metallic glasses, where deformation processes like shear transformation zones are governed by local descriptors such as the extent of structural ordering, compositional heterogeneities, etc. [10-13]. However, metals with amorphous regions confined to the grain boundary possess additional mechanistic considerations such as the absorption/emission of defects from and to the adjoining crystals, and additional local descriptors such as the thickness and chemical or structural gradient layering of the grain boundary region. Micromechanical tests are especially useful for studying the effects of grain boundary descriptors in detail, as plastic deformation can be rapidly probed. For example, Liu et al. used microcompression testing to isolate the interfacial shear stress between amorphous ZrCu and crystalline Zr nanolayers, finding that the amorphous layers were capable of accommodating plastic strains of at least 18% through interfacial sliding [14]. Su et al. found that dopant-rich, thick (~5 nm) grain boundaries enabled ultrahigh yield strengths above 4 GPa and uniform compressive strain up to 20% in nanostructured intermetallics [15]. Zhuang et al. fabricated a multi-principal element alloy thin film incorporating amorphous nanodomains and found increased strength compared to either a purely nanocrystalline thin film or a purely amorphous thin film [16], showing that amorphous and ordered domains can be combined to improve properties.

Although amorphous complexions have been shown to improve strength [17] and ductility [18] in nanocrystalline Cu-Zr, the effect of complexion descriptors such as thickness on mechanical deformation has yet to be studied. Prior simulations [19] have predicted that thicker complexions may be able to spread out the plastic strain over a larger region, suggesting improved resistance to premature failure. In this work, two model nanocrystalline samples are created, where all microstructural descriptors are the same except for complexion thickness, as controlled by the cooling rate following a heat treatment. *In-situ* scanning electron microscopy (SEM) microcompression testing is then performed to study the effect of complexion thickness on plasticity to different deformation levels. The results of this study demonstrate that thicker amorphous complexions lead to increased plasticity and failure resistance in nanocrystalline metal alloys, informing future design of alloys and their processing routes.

Cu-3 at.% Zr (denoted simply as "Cu-3Zr" hereafter) alloys were prepared from elemental powders of Cu (Alfa Aesar, 99%, -325 mesh) and Zr (Atlantic Equipment Engineers, 99.5%, -20+60 mesh) that were mechanically alloyed for 10 h in a SPEX SamplePrep 8000M high-energy ball mill to refine grains and induce chemical mixing. Hardened steel vials and milling media were used, with milling occurring inside a glovebox under Ar with < 0.15 ppm $O_2$ levels and 1 wt.% stearic acid (Alfa Aesar) added to minimize cold welding. Approximately 9 grams of alloyed powder were consolidated first for 10 min under 25 MPa at room temperature inside the glovebox to form green bodies (cylinders with ~1 cm height, ~1.4 cm diameter), followed by sintering for 1 h under 50 MPa at 950 °C in a vacuum hot press. The consolidated pellets were annealed at 950 °C for 5 min and quenched on a liquid nitrogen-cooled Al heat sink to preserve the high-temperature grain boundary structures and create two distinct sets of samples: (1) a *fast-quenched* sample taken from the pellet face in direct contact with the heat sink and (2) a *slow-cooled* sample from the pellet face far away from the heat sink. Additional details about the general quenching setup and expected cooling rates can be found in Ref. [20] where the cooling rates differed by more than two orders of magnitude for the opposite faces of the pellet, with resulting complexion distributions discussed below. Sample surfaces were polished with SiC grinding paper down to 1200 grit before microstructural characterization. X-ray diffraction (XRD) scans were conducted using a Rigaku Ultima III X-ray diffractometer with a Cu Kα radiation source operated at 40 kV and 30 mA, and a one-dimensional D/teX Ultra detector. Phase identification and fraction were obtained using an integrated powder X-ray analysis software package (Rigaku PDXL) and Rietveld refinements were performed with PDXL2 analysis software. Transmission electron microscopy (TEM) specimens from the fast-quenched and slow-quenched regions were prepared using the lift-out method with the FEI Quanta 3D equipped with a $Ga^+$ ion beam and OmniProbe, with the sample welded to a Mo grid and further thinning using a Tescan GAIA3 SEM/FIB. Scanning transmission electron microscopy (STEM) was used to identify and measure amorphous complexions using a JEOL JEM-ARM300F Grand ARM TEM with double Cs correctors operated at 300 kV.

Micropillars were fabricated using two FIB methods: (1) annular milling and (2) lathe milling. In both methods, pillars with a height-to-diameter ratio of ~2:1 were fabricated to prevent plastic buckling [21] and the diameters ranged from ~3-5 μm, multiple orders of magnitude larger than the grain size to avoid size effects. The annular pillars were milled using a FEI Helios 600 dual-beam $Ga^+$ SEM/FIB and can be made quickly, allowing for a large number of samples to be tested in rapid succession to relatively large applied strains. The lathed pillars were milled to be taper-free using a FEI Quanta 3D FEG dual-beam $Ga^+$ SEM/FIB, which is significantly more time consuming, resulting in fewer pillars that were used here to test to smaller plastic strains and determine any potential effects of the pillar geometry. Details about how the lathing steps were adapted from the approach first developed by Uchic and Dimiduk [21] can be found in Ref. [22]. The *in-situ* compression tests were performed using a FemtoTools nanomechanical testing system (Model FT-NMT03 for the lathe milled pillars and Model FT-NMT04 for the annular milled pillars) under SEM observation. The load was applied by a flat platen with a cross section of either 5 μm × 5 μm (annular) or 20 μm × 25 μm (lathed). All tests were conducted in displacement-control mode using a subnanometer-resolution piezo-based actuation system and nominal strain rates of $10^{-1}$-$10^{-3}$ $s^{-1}$ were applied.

The microstructures of both the slow-cooled and fast-quenched samples are shown through representative bright field (BF)-STEM images in Figs. 1(a) and (b). Grain sizes were measured to be 44 ± 14 nm and 47 ± 17 nm for the slow-cooled and fast-quenched samples, respectively, (i.e., measurements were taken from opposite sides of the pellet) and are in reasonable agreement with the XRD measurement, which gave a value of 36 nm (Fig. 1(c)). The XRD scan is not targeted at any specific location but rather at the general cross-section. TEM grain size was calculated by measuring at least 65 grains from each sample type, by identifying grains that were strongly diffracting and calculating the average circular equivalent diameter. Additional TEM images and grain size distributions are presented in Figures S1 and S2, respectively, in the Supplementary Materials. Only trace amounts of ZrC impurity phases were observed, with no differences between the two samples. These measurements confirm that grain size is unaffected by the difference in cooling rate between the two samples. The time at elevated temperature is nearly identical for the two samples, with the differences in cooling rate only resulting in changes to time-at-temperature of the order of seconds [20]. Higher magnification BF-STEM images were collected to identify amorphous complexions and measure their thickness, with examples shown in Figs. 1(d) and (f). At least a dozen amorphous complexions were measured (distributions shown in Fig. 1(e)), with the average thickness from the fast-quenched samples (3.24 ± 1.15 nm) being ~1.6× the average thickness from the slow-cooled samples (1.99 ± 1.16 nm). These complexions are distributed throughout the grain boundary network, with atomistic models predicting that a significant fraction of boundaries but not all have transformed [23]. The effect of cooling rates of Cu alloys on the matrix grain size and complexion thickness has been measured previously [20], so these distributions are confirming a known behavior here. From here on, the slow-cooled sample is referred to as having *thinner complexions*, while the fast-quenched sample is referred to as having *thicker complexions*. We emphasize that the grain size, global Zr concentration, segregation state (see Ref. [20]), and carbide content are the same between the two samples, so the only difference between the two sets of materials is the amorphous complexion thickness.

.

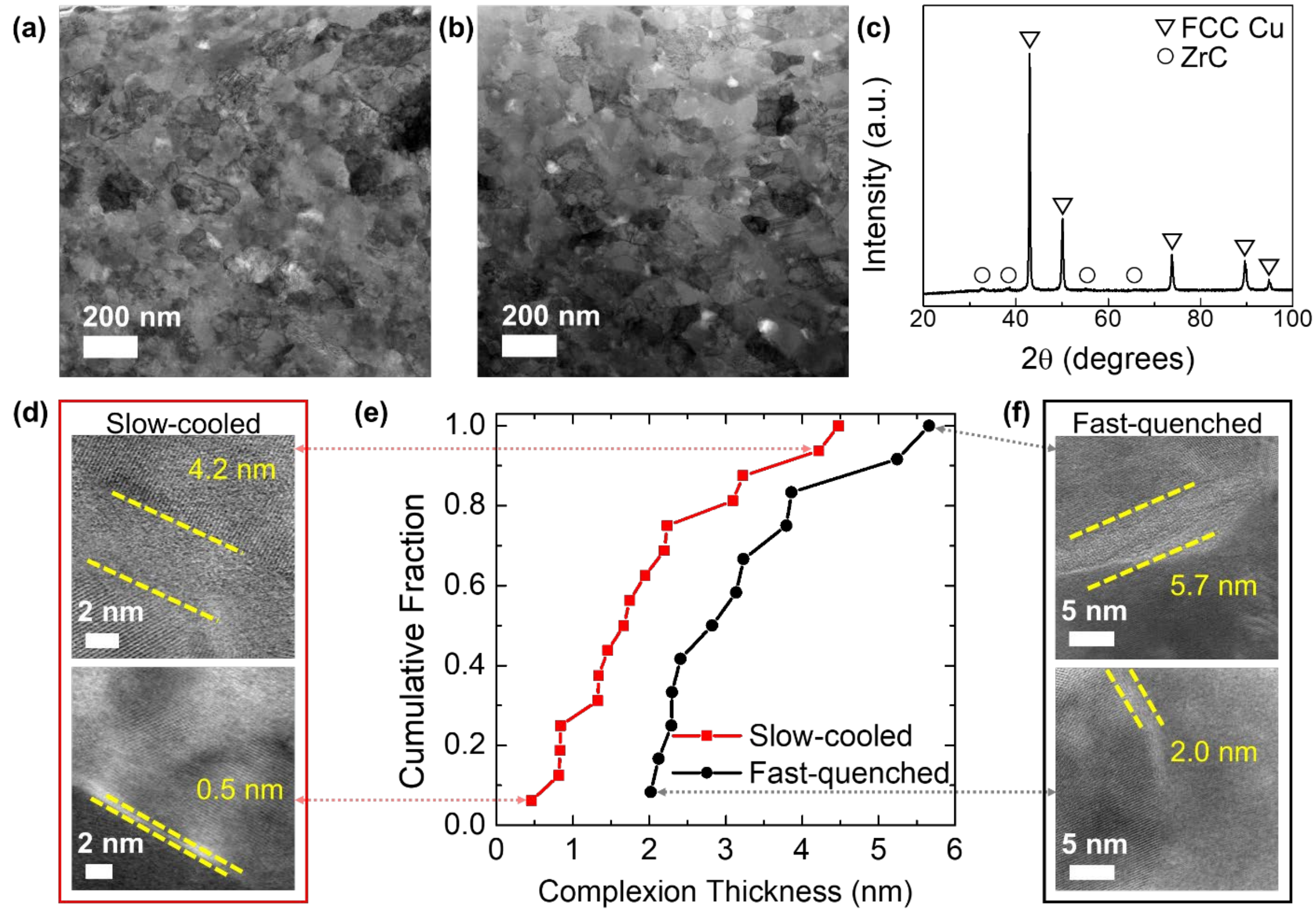

**Fig. 1. Low magnification bright field (BF)-STEM images from the (a) slow-cooled and (b) fast-quenched samples. (c) XRD scan showing a primary FCC Cu phase, with minority impurity ZrC phase. BF-STEM images of examples of amorphous complexions outlined with yellow dashed lines from the (d) slow-cooled sample and (f) the fast-quenched sample. (e) Amorphous complexion thicknesses analyzed from STEM data for the slow-cooled and fast-quenched faces, showing that the fast-quenched sample has thicker amorphous complexions.**

Representative pillars from each of the two milling procedures, lathe-milled and annular-milled, are shown in Fig. 2(a) and (b), respectively, prior to deformation. The pillars are imaged with the SEM stage tilted to 52° for comparison. Fig. 2(c) shows an example of a lathe-milled pillar imaged head-on without stage tilt, showing the lack of taper. Lathe-milled pillars were tested at a strain rate of $10^{-3}$ $s^{-1}$, and three different strain rates in the range $10^{-3}$-$10^{-1}$ $s^{-1}$ strain rates were used for the annular pillars. One representative engineering stress-strain curve for each pillar type from the sample with thicker complexions tested at $10^{-3}$ $s^{-1}$ is plotted in Fig. 2(d). The annular-milled pillars were compressed to relatively large plastic strains, to obtain statistical information on preferred failure mode and explore heavy deformed conditions. The lathe-milled pillars were tested to relatively small plastic strains, to probe the early stages of plasticity and enable observations of plastic morphology associated with yielding.

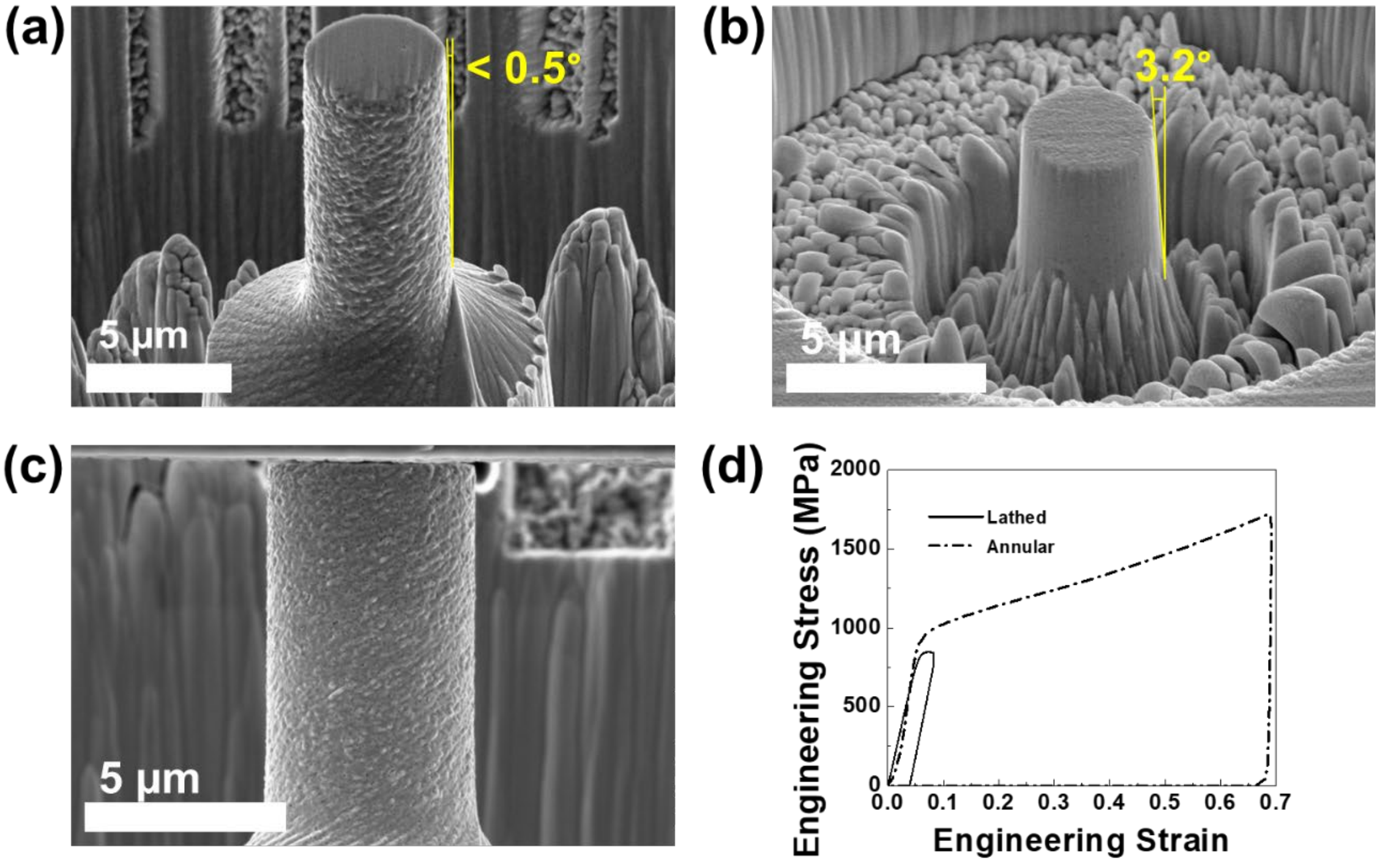

**Fig. 2. Micropillars were fabricated using two methods, with an example of a (a) lathe-milled and (b) annular-milled pillar imaged with the SEM stage tilted to 52°. (c) A lathe-milled pillar, imaged without stage tilt. (d) Representative engineering stress-strain curves from the sample with thicker complexions for annular and lathed pillars tested at $10^{-3}$ $s^{-1}$.**

For the first set of tests, 45 annular pillars were compressed to large plastic strains of >35%. Examples of deformed pillars after compression at the $10^{-3}$ $s^{-1}$ strain rate with thinner and thicker complexions are shown in Fig. 3(a) and (b). Two different failure modes of localized shearing and homogeneous/uniform barreling were observed. Most pillars showed some degree of stable plastic deformation, indicated with the yellow dashed lines, with the second pillar in Fig. 3(a) being the exception which appears to rapidly shear band; in this pillar, strain localization occurs before stable plastic flow can be developed and there is no curvature of the pillar from barreling. In nanocrystalline metals, where the grain boundaries facilitate plastic deformation through mechanisms such as grain boundary sliding [24, 25], compression can induce strain localization that occurs as shear banding spanning the micropillar diameter, sometimes resembling the deformation of metallic glasses [26, 27]. It is unlikely that the observed shear band on the micropillars formed along a single grain boundary path, as most pillars showed some plastic deformation before localization. Micropillars of nanocrystalline Ni-W also experienced plastic deformation before catastrophic localization, which the authors attributed to grains limiting shear band formation [28]. Further, Balbus et al. showed that the formation of amorphous complexions in nanocrystalline Al alloys suppress shear localization [29].

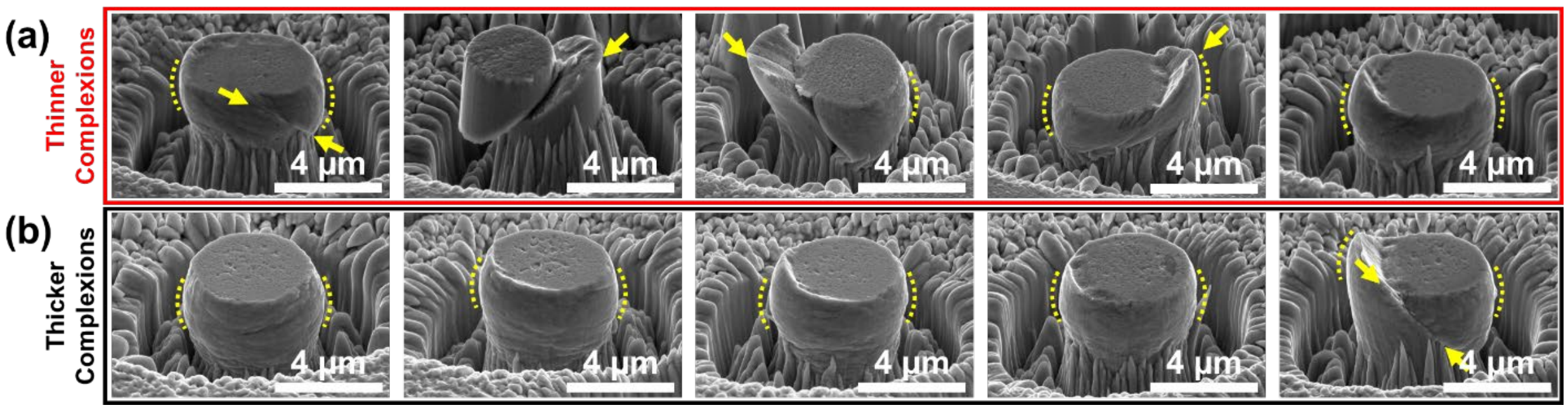

**Fig. 3. Heavily deformed annular-milled pillars with (a) thinner and (b) thicker complexions. Yellow arrows highlight where localized plastic deformation has occurred in the form of shear banding and yellow dashed lines outline the curvature associated with stable plastic deformation.**

Plastic localization is more commonly observed in the sample with thinner complexions, with statistical evidence presented in Fig. 4. Fig. 4(a) shows a representative example of "localized deformation," where obvious shear banding is observed. Any pillar shown in Fig. 3 that has slip localization denoted by a yellow arrow would count in this category. Fig. 4(b) shows a representative example of "homogenous deformation," where no localization is observed and instead the pillars only barrel outwards during compression. Only those pillars in Fig. 3 without yellow arrows would fit this classification. Fig. 4(c) shows all of the annular-milled pillars across the two sample types and three different strain rates. Some pillars from each sample set experience each of the failure modes, yet the thicker complexion material has less plastic localization overall. While differences in the percentages of pillars that fail through a given mode can be seen in Fig. 4(c), there is no consistent trend across the strain rates probed where increasing strain rate increases the tendency for one failure mode over the other. We acknowledge that the range of strain rates tested was relatively small and that all strain rates are relatively low, so more thorough investigations may be warranted in the future. However, as there is no obvious trend in our data, Fig. 4(d) combines all the annular pillar data to allow for a comparison using the full dataset. The sample with thicker complexions is significantly less likely to experience plastic strain localization (e.g. shear band through pillar diameter), instead deforming through a homogeneous deformation mode. Increased plastic localization for the sample with thinner complexions is in reasonable agreement with prior literature, considering that thinner complexions have an overall lower amorphous grain boundary volume compared to thicker complexions. In nanocrystalline Ni-W, achievement of a more ordered grain boundary state through thermal relaxation led to more localization [28]. For nanostructured crystalline-amorphous composites, several groups have reported that thicker grain boundaries lead to higher strengths [7, 8, 30]. Notably, Qian et al.

simulated nanocrystalline Cu with a grain size of 12.5 nm and amorphous Ta-doped grain boundaries that have thicknesses varying from 0.5 to 16 nm, finding that the strength increases as the amorphous boundary thickness increases until a critical threshold of 4 nm is reached and strength begins to decrease [31]. Below the critical thickness, dislocation-related plasticity dominates, while above the critical thickness shear-transformation zones dominate contributions to the overall strength. At the critical thickness, the contributions from dislocation activity from the abutting crystals and the shear transformation zones in the amorphous boundary were nearly equal [31]. Qian et al. also found that as the amorphous boundary thickness reaches 10s to 100s of nm (reaching and exceeding the matrix grain size by an order of magnitude), deformation is dominated by shear banding. This means that is amorphous complexions became thick enough that a shear band could operate through the specimen thickness uninterrupted, localized failure would be expected. However, even the thickest complexion measured for the nanocrystalline Cu-Zr in this work is below 10 nm and thus should have only dislocation-shear transformation zone interactions. In addition, not all boundaries have transformed to amorphous complexions, so no continuous amorphous network exists.

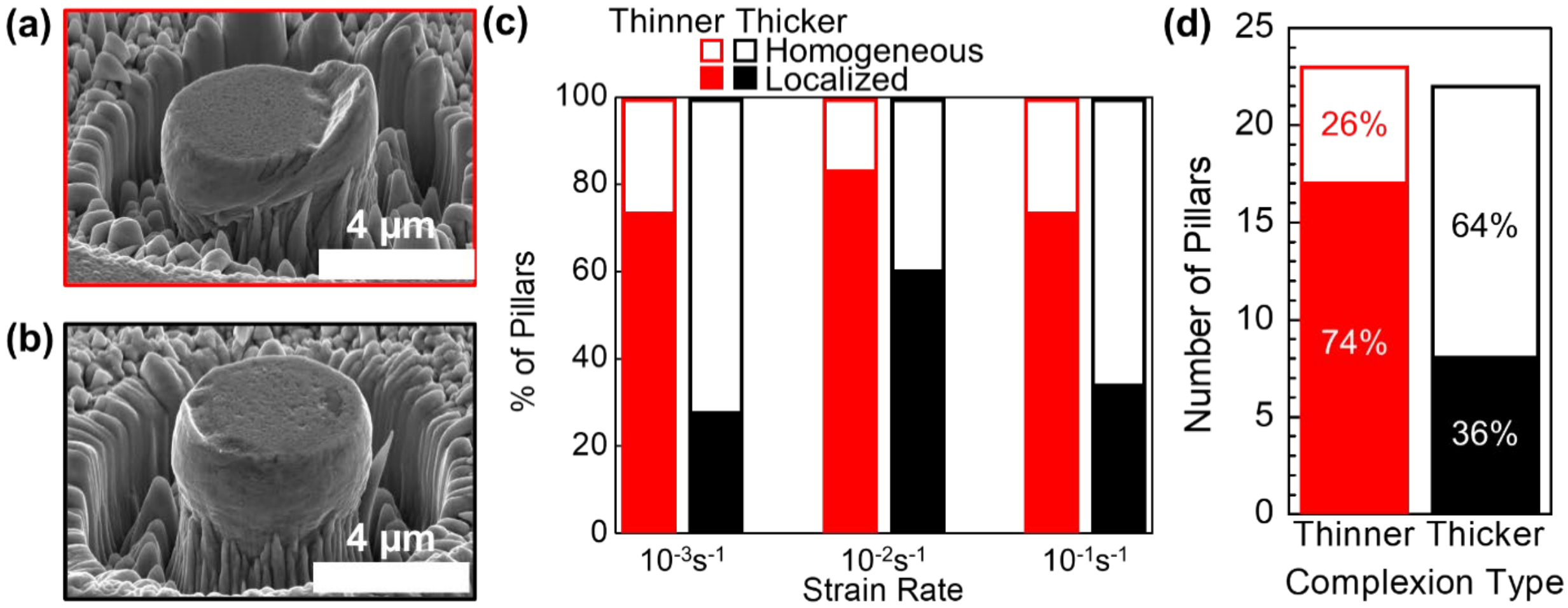

**Fig. 4. Representative examples of (a) localized and (b) homogeneous deformation modes. Bar charts of the (c) percentage of pillars with thinner and thicker complexions that deform by either homogeneous or localized**

**modes at three strain rates and (d) the total number of pillars with each complexion type and deformation mode.**

The annular pillars deformed to large plastic strains show the severe deformation case, so lathe-milled pillars were next studied to understand differences in early plastic flow. Six pillars of each sample type were compressed to strains of <5%. Examples of deformed pillars are shown in Figs. 5(a) and (b) for samples with thinner and thicker complexions, respectively. The early stages of plastic deformation are more subtle, yet our annular pillar experiments tell us that we are looking for signs of localization or homogeneous deformation. Thus, using lathe milling to achieve a taper-free geometry and *in-situ* imaging to allows for more direct quantification of plasticity. To quantify any differences, we define *plastic asymmetry* as:

$$plastic\ asymmetry = \frac{|area_L - area_R|}{total\ area} \quad (1)$$

where $area_L$ is the area of the left half of the pillar and $area_R$ is the area of the right half of the pillar. Figs. 5(a) and 5(b) show the absolute centerline of the pillars with dashed yellow lines. A plastic asymmetry value close to 0 indicates uniform plastic deformation in a homogeneous mode. In contrast, a larger value means that one side of the pillar is deforming much more than the other, signaling the early stages of strain localization. The results for thinner and thicker complexions are summarized in Fig. 5(c). The sample with thicker complexions has an average plastic asymmetry that is lower by a factor of 2× as compared to the thinner complexion sample.

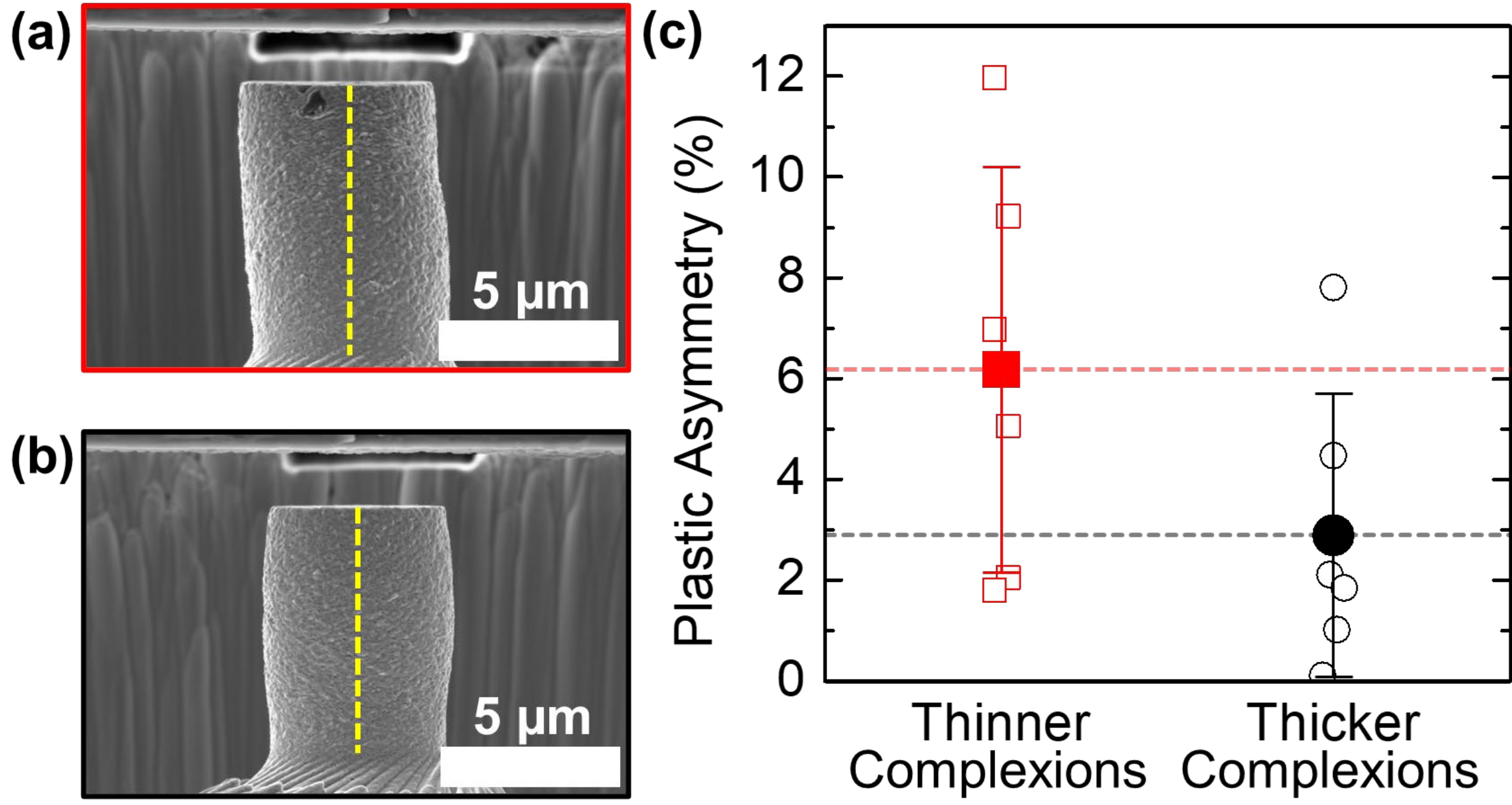

**Fig. 5. Representative examples of pillar in early deformation stages from (a) the sample with thinner complexions that displays more localized deformation and (b) the sample with thicker complexions, which displays more uniform barreling. (c) Compiled results for plastic asymmetry showing that the material with the thicker complexions has much lower plastic asymmetry, signaling more uniform plasticity. The open data points correspond to individual pillars, the filled data points and dashed lines indicate the average plastic asymmetry values, and the error bars show one standard deviation.**

Amorphous complexions could hypothetically degrade plasticity if they acted as a soft matrix while the grains acted as harder precipitates, as is sometimes observed in metallic glasses. As complexion thickness increased, the volume fraction of softer matrix would also increase and eventually create a percolating pathway for a localization event. However, our results here disprove such a scenario, as both the annular-milled and lathe-milled pillars consistently show that nanocrystalline alloys with thicker amorphous complexions experience more homogeneous plasticity. The current study presents the first experimental evidence that thicker amorphous complexions act as beneficial features for plasticity. Complexion thickness has been discussed as a descriptor leading to higher strengths, with dislocation interactions playing an important role. Phan et al. studied an amorphous-crystalline interface to determine how incoming dislocations are accommodated in a bilayer Cu and glassy Cu-Zr sample, finding a multi-stage deformation process where early incoming dislocations caused localized plastic strains at the interface and formed new shear transformation zones [32]. With increased loading, the shear transformation zones coalesce as tens of dislocations were absorbed into the glassy layer [32]. As mentioned earlier, Qian et al. determined a critical amorphous grain boundary thickness in simulated Cu-CuTa where shear transformation zones dominate the deformation [31]. Imaging cross sections of deformed pillars of nanocrystalline Ni that displayed large plasticity, Wei et al. found dislocations distributed along the amorphous-crystal interfaces and attributed plastic co-deformation between grains and grain boundaries to this behavior [33]. Dislocations can be impeded or trapped at amorphous boundaries, which prevents the formation of localized shear bands across multiple grains, thus enabling stable plastic deformation for the pillars with thicker complexions here. Finally, Pan and Rupert simulated a Cu-Zr alloy with amorphous intergranular films to study dislocation accommodation and crack nucleation, finding that thicker films could more effectively absorb dislocations by spreading the plastic zone out within the complexion to slow the nucleation and growth of cracks [19]. Based on this supporting experimental and computational evidence, thicker amorphous complexions appear to suppress localization by absorbing more dislocation content from the adjoining crystals prior to failure, enabling larger homogeneous plastic strains.

In the present study, the plastic flow and failure of nanocrystalline Cu-Zr alloys having either thinner or thicker amorphous complexions was investigated using in-situ SEM micropillar compression testing. The sample with thinner complexions experienced stronger non-uniform plastic deformation and localized plastic flow when pushed to large deformations, with early signs of localization also observed at small plastic strains. In contrast, the sample with thicker complexions (~1.6× the average thickness from the slow-cooled samples) displayed increased plasticity and higher damage tolerance, with more uniform plastic deformation in both small and large strain conditions. Heavily deformed annular pillars with thinner complexions experienced localized deformation 74% of the time compared to pillars with thicker complexions that experienced localized deformation only 36% of the time. Early deformation stages probed in lathed pillars showed that the sample with thinner complexions had an average plastic asymmetry that is about 2× larger compared to the thicker complexion sample. In general, this work clearly identifies amorphous complexion thickness as an important descriptor for determining the plasticity of nanocrystalline metals. Increasing the thickness of the disordered regions between grains increases the ability to diffuse strain concentrations during the absorption of dislocations [34] and leading to toughening [19, 35], as evidenced by atomistic modeling studies in the literature.

**Declaration of Interest Statement**

The corresponding author, Timothy J. Rupert, is an Editor for Scripta Materialia and was not involved in the editorial review or the decision to publish this article.

**Acknowledgements**

This work was supported by the U.S. Department of Energy, Office of Science, Basic Energy Sciences, under Award No. DE-SC0025195. E.C.H. received additional support from the UCI-LANL-SoCal Hub Graduate Student Research Fellowship Program. This work was performed, in part, at the Center for Integrated Nanotechnologies, an Office of Science User Facility operated for the U.S. Department of Energy (DOE) Office of Science. Los Alamos National Laboratory, an affirmative action equal opportunity employer, is managed by Triad National Security, LLC for the U.S. Department of Energy's NNSA, under contract 89233218CNA000001. Sandia National Laboratories is a multimission laboratory managed and operated by National Technology & Engineering Solutions of Sandia, LLC, a wholly owned subsidiary of Honeywell International, Inc., for the U.S. DOE's National Nuclear Security Administration under contract DE-NA-0003525. The views expressed in the article do not necessarily represent the views of the U.S. DOE or the United States Government. The authors acknowledge support by Jonathan Gigax for access to the FemtoTools NMT-04. The research reported here made use of the shared facilities of the Materials Research Science and Engineering Center (MRSEC) at UC Santa Barbara: NSF DMR–2308708. The authors acknowledge the use of facilities and instrumentation at the UC Irvine Materials Research Institute (IMRI), which is supported in part by the National Science Foundation through the UC Irvine Materials Research Science and Engineering Center (DMR-2011967). SEM, FIB, and EDS work was performed using instrumentation funded in part by the National Science Foundation Center for Chemistry at the Space-Time Limit (CHE-0802913).

## References

[1] M.A. Meyers, A. Mishra, D.J. Benson, Mechanical properties of nanocrystalline materials, Progress in Materials Science 51(4) (2006) 427-556.
[2] M. Dao, L. Lu, R. Asaro, J.T.M. De Hosson, E. Ma, Toward a quantitative understanding of mechanical behavior of nanocrystalline metals, Acta Materialia 55(12) (2007) 4041-4065.
[3] N. Hansen, Polycrystalline strengthening, Metallurgical Transactions A 16(12) (1985) 2167-2190.
[4] A. Lasalmonie, J. Strudel, Influence of grain size on the mechanical behaviour of some high strength materials, Journal of Materials Science 21(6) (1986) 1837-1852.
[5] K. Xu, X. Sheng, A. Mathew, E. Flores, H. Wang, Y. Kulkarni, X. Zhang, Mechanical Behavior and Thermal Stability of Nanocrystalline Metallic Materials with Thick Grain Boundaries, JOM 76(6) (2024) 2914-2928.
[6] Q. An, Z. Yan, L. Bai, S. Zheng, Achieving superior matching of strength, plasticity, and strain hardening in multilayers by introducing metastable amorphous interface phase, Scripta Materialia 252 (2024) 116258.
[7] J. Ding, D. Neffati, Q. Li, R. Su, J. Li, S. Xue, Z. Shang, Y. Zhang, H. Wang, Y. Kulkarni, Thick grain boundary induced strengthening in nanocrystalline Ni alloy, Nanoscale 11(48) (2019) 23449-23458.
[8] G. Wu, K.-C. Chan, L. Zhu, L. Sun, J. Lu, Dual-phase nanostructuring as a route to high-strength magnesium alloys, Nature 545(7652) (2017) 80-83.
[9] K. Madhav Reddy, J. Guo, Y. Shinoda, T. Fujita, A. Hirata, J. Singh, J.W. McCauley, M. Chen, Enhanced mechanical properties of nanocrystalline boron carbide by nanoporosity and interface phases, Nature Communications 3(1) (2012) 1052.
[10] M.M. Trexler, N.N. Thadhani, Mechanical properties of bulk metallic glasses, Progress in Materials Science 55(8) (2010) 759-839.
[11] A. Greer, Y. Cheng, E. Ma, Shear bands in metallic glasses, Materials Science and Engineering: R: Reports 74(4) (2013) 71-132.
[12] J.R. Greer, J.T.M. De Hosson, Plasticity in small-sized metallic systems: Intrinsic versus extrinsic size effect, Progress in Materials Science 56(6) (2011) 654-724.
[13] T.C. Hufnagel, C.A. Schuh, M.L. Falk, Deformation of metallic glasses: Recent developments in theory, simulations, and experiments, Acta Materialia 109 (2016) 375-393.
[14] M. Liu, J. Huang, Y. Fong, S. Ju, X. Du, H. Pei, T. Nieh, Assessing the interfacial strength of an amorphous–crystalline interface, Acta Materialia 61(9) (2013) 3304-3313.
[15] R. Su, D. Neffati, J. Cho, Z. Shang, Y. Zhang, J. Ding, Q. Li, S. Xue, H. Wang, Y. Kulkarni, High-strength nanocrystalline intermetallics with room temperature deformability enabled by nanometer thick grain boundaries, Science Advances 7(27) (2021) eabc8288.
[16] Q. Zhuang, D. Liang, J. Luo, K. Chu, K. Yan, L. Yang, C. Wei, F. Jiang, Z. Li, F. Ren, Dual-Nano composite design with grain boundary segregation for enhanced strength and plasticity in CoCrNi-CuZr thin films, Nano Letters 25(2) (2024) 691-698.
[17] A. Khalajhedayati, Z. Pan, T.J. Rupert, Manipulating the interfacial structure of nanomaterials to achieve a unique combination of strength and ductility, Nature Communications 7(1) (2016) 10802.
[18] J.L. Wardini, C.M. Grigorian, T.J. Rupert, Amorphous complexions alter the tensile failure of nanocrystalline Cu-Zr alloys, Materialia 17 (2021) 101134.

[19] Z. Pan, T.J. Rupert, Amorphous intergranular films as toughening structural features, Acta Materialia 89 (2015) 205-214.
[20] C.M. Grigorian, T.J. Rupert, Critical cooling rates for amorphous-to-ordered complexion transitions in Cu-rich nanocrystalline alloys, Acta Materialia 206 (2021) 116650.
[21] M.D. Uchic, D.M. Dimiduk, A methodology to investigate size scale effects in crystalline plasticity using uniaxial compression testing, Materials Science and Engineering: A 400 (2005) 268-278.
[22] T. Lei, E.C. Hessong, J. Shin, D.S. Gianola, T.J. Rupert, Intermetallic particle heterogeneity controls shear localization in high-strength nanostructured Al alloys, Acta Materialia 240 (2022) 118347.
[23] P. Garg, Z. Pan, V. Turlo, T.J. Rupert, Segregation competition and complexion coexistence within a polycrystalline grain boundary network, Acta Materialia 218 (2021) 117213.
[24] E.N. Borodin, A.E. Mayer, A simple mechanical model for grain boundary sliding in nanocrystalline metals, Materials Science and Engineering: A 532 (2012) 245-248.
[25] H. Van Swygenhoven, P. Derlet, Grain-boundary sliding in nanocrystalline fcc metals, Physical review B 64(22) (2001) 224105.
[26] E.R. Homer, Examining the initial stages of shear localization in amorphous metals, Acta Materialia 63 (2014) 44-53.
[27] A. Hodge, T. Furnish, A. Navid, T. Barbee Jr, Shear band formation and ductility in nanotwinned Cu, Scripta Materialia 65(11) (2011) 1006-1009.
[28] A. Khalajhedayati, T.J. Rupert, Emergence of localized plasticity and failure through shear banding during microcompression of a nanocrystalline alloy, Acta materialia 65 (2014) 326-337.
[29] G.H. Balbus, F. Wang, D.S. Gianola, Suppression of shear localization in nanocrystalline Al–Ni–Ce via segregation engineering, Acta Materialia 188 (2020) 63-78.
[30] G. Wu, S. Balachandran, B. Gault, W. Xia, C. Liu, Z. Rao, Y. Wei, S. Liu, J. Lu, M. Herbig, W. Lu, G. Dehm, Z. Li, D. Raabe, Crystal–glass high-entropy nanocomposites with near theoretical compressive strength and large deformability, Advanced Materials 32(34) (2020) 2002619.
[31] L. Qian, W. Yang, J. Luo, Y. Wang, K. Chan, X.-S. Yang, Amorphous thickness-dependent strengthening–softening transition in crystalline–amorphous nanocomposites, Nano Letters 23(23) (2023) 11288-11296.
[32] T. Phan, J. Rigelesaiyin, Y. Chen, A. Bastawros, L. Xiong, Metallic glass instability induced by the continuous dislocation absorption at an amorphous/crystalline interface, Acta Materialia 189 (2020) 10-24.
[33] B. Wei, W. Wu, D. Xie, M. Nastasi, J. Wang, Strength, plasticity, thermal stability and strain rate sensitivity of nanograined nickel with amorphous ceramic grain boundaries, Acta Materialia 212 (2021) 116918.
[34] Y. Piao, K.C. Le, Thermodynamic theory of dislocation/grain boundary interaction, Continuum Mechanics and Thermodynamics 34(3) (2022) 763-780.
[35] P. Garg, T.J. Rupert, Local structural ordering determines the mechanical damage tolerance of amorphous grain boundary complexions, Scripta Materialia 237 (2023) 115712.

**Supplementary Materials**

**Note 1. Additional STEM images of the microstructure.**

Figure S1 presents the bright field (BF)- and high angle annular dark field (HAADF)-STEM images from lamella lifted out from opposite areas of the pellet cross section, representing the slow-cooled and fast-quenched samples.

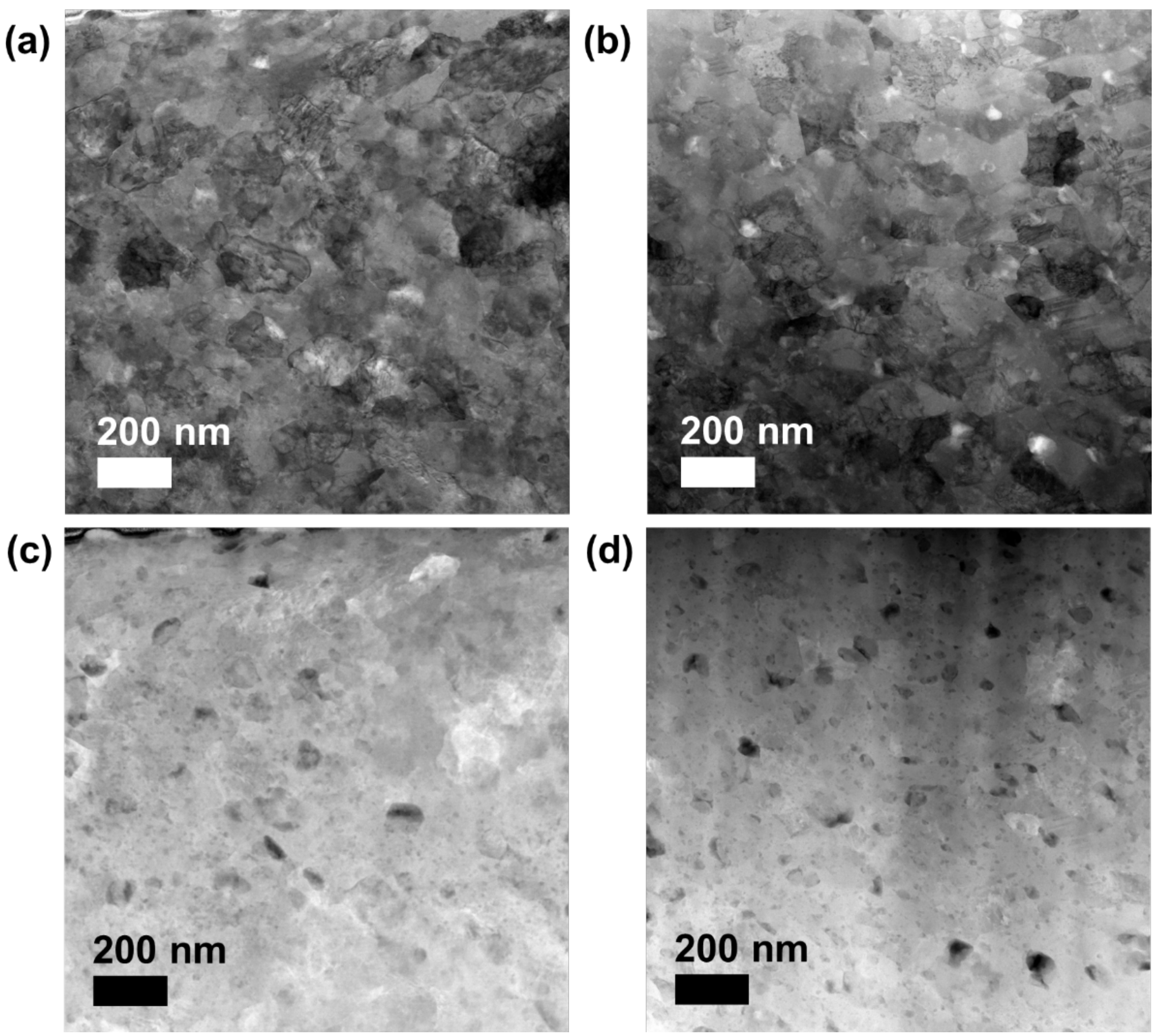

**Fig. S1. Low magnification BF-STEM images from the (a) slow-cooled and (b) fast-quenched samples, and HAADF-STEM images from the (c) slow-cooled and (d) fast-quenched samples.**

## Supplementary Note 2

### Note 2. Grain size distribution.

Figure S2 presents the distribution of grain sizes measured from the slow-cooled and fast-quenched areas of the bulk pellet face. The grain sizes are approximately the same across the two samples.

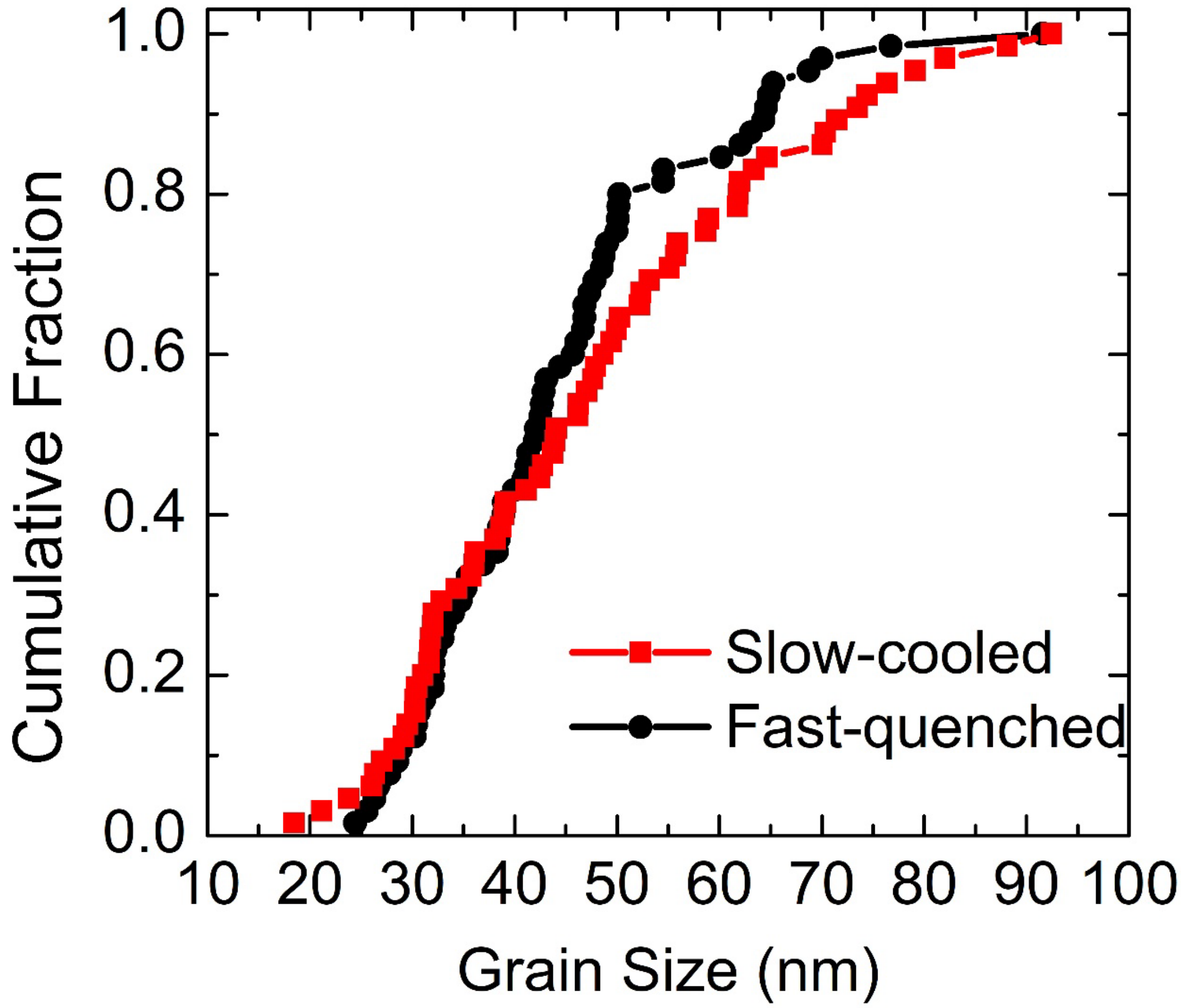

**Fig. S2. Grain size distribution of 65 grains each from the slow-cooled (red square) and fast quenched (black circle).**

## Supplementary Note 3

## Note 3. Engineering stress-strain data.

Figure S3 presents the engineering stress-strain curves from all of the tests, with the annular pillar tests shown in Fig. S3(b) and then separated in (c)-(e) to more clearly show each strain rate.

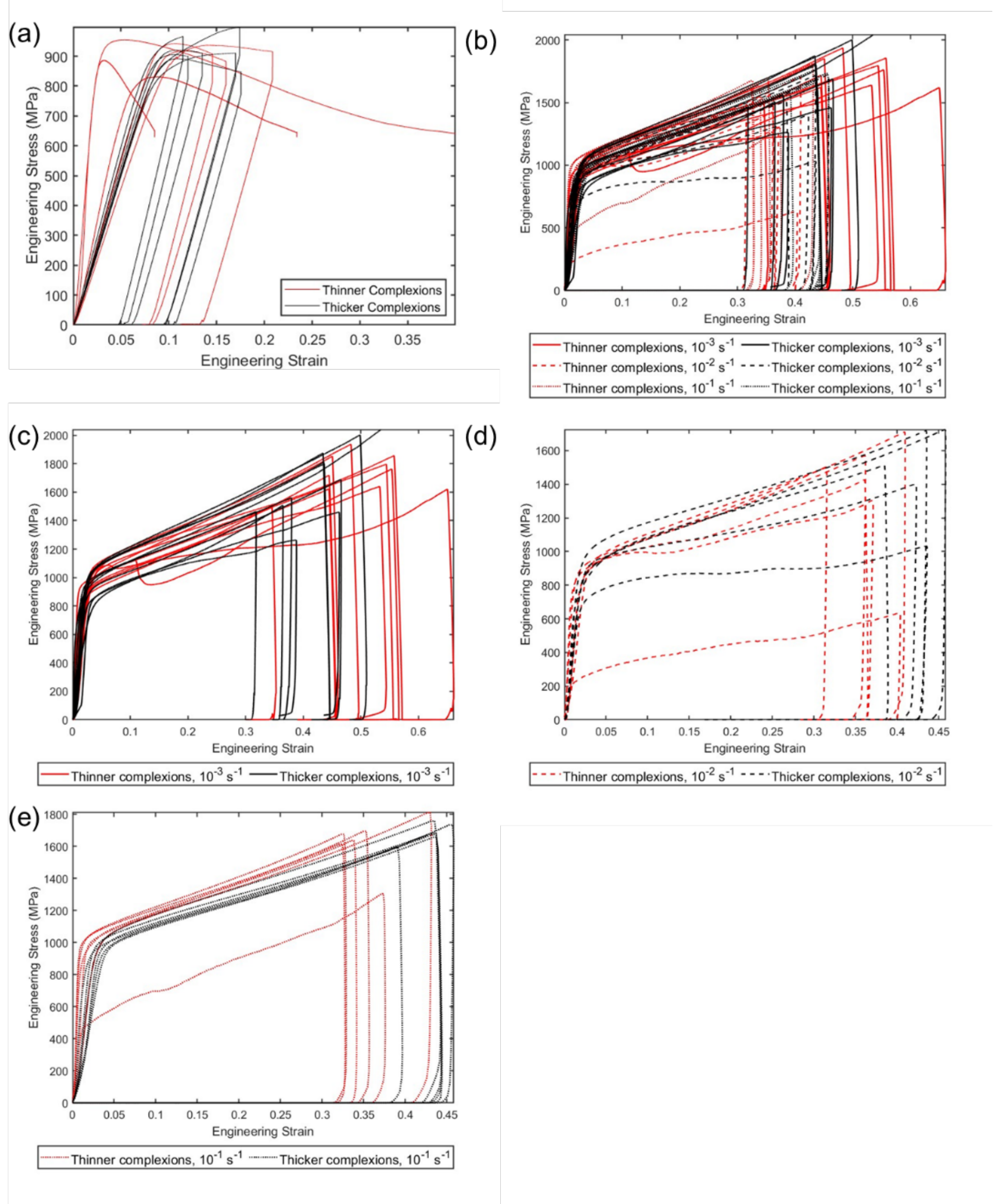

**Fig. S3. Engineering stress-strain curves for (a) lathed pillars, (b) annular pillars tested across all strain rates, and annular pillars tested with a strain rate of (c) $10^{-3}$ $s^{-1}$, (d) $10^{-2}$ $s^{-1}$, and (e) $10^{-1}$ $s^{-1}$.**